\documentclass[10pt,twocolumn,letterpaper]{article}

\usepackage{iccv}
\usepackage{times}
\usepackage{epsfig}
\usepackage{graphicx}
\usepackage{amsmath}
\usepackage{amssymb}
\usepackage{subcaption}
\usepackage{booktabs}
\usepackage{verbatim}
\usepackage{textcomp}
\usepackage{float}
\usepackage{balance}
\usepackage{multirow}
\usepackage{rotating}
\usepackage{arydshln}

\definecolor{forestgreen}{rgb}{0.13, 0.55, 0.13}

\newcommand{\mypara}[1]{\vspace{1mm}\noindent\textbf{#1}}

\usepackage[breaklinks=true,bookmarks=false]{hyperref}

\iccvfinalcopy 

\def\iccvPaperID{****} 

\begin{document}

\title{How Can We Tame the Long-Tail of Chest X-ray Datasets?}

\author{Arsh Verma\\
Wadhwani Institute for Artificial Intelligence (Wadhwani AI), India\\
{\tt\small arsh@wadhwaniai.org}}
\maketitle

\begin{abstract}
    Chest X-rays (CXRs) are a medical imaging modality that is used to infer a large number of abnormalities. While it is hard to define an exhaustive list of these abnormalities, which may co-occur on a chest X-ray, few of them are quite commonly observed and are abundantly represented in CXR datasets used to train deep learning models for automated inference. However, it is challenging for current models to learn independent discriminatory features for labels that are rare but may be of high significance. Prior works focus on the combination of multi-label and long tail problems by introducing novel loss functions or some mechanism of re-sampling or re-weighting the data. Instead, we propose that it is possible to achieve significant performance gains merely by choosing an initialization for a model that is closer to the domain of the target dataset. This method can complement the techniques proposed in existing literature, and can easily be scaled to new labels. Finally, we also examine the veracity of synthetically generated data to augment the tail labels and analyse its contribution to improving model performance. 
\end{abstract}

\section{Introduction}
\label{sec:introduction}

Chest X-rays (CXRs) are used to infer a large number of abnormalities by radiologists across the world. Most of these are routine findings and medical professionals may see multiple instances of those in each batch of patients they attend to. However, there are many abnormalities which are encountered quite rarely \cite{zhou2021review}, to the extent that even specialists may have studied them only in theory. 

\begin{figure}[t]
    \centering
    \includegraphics[width=\linewidth]{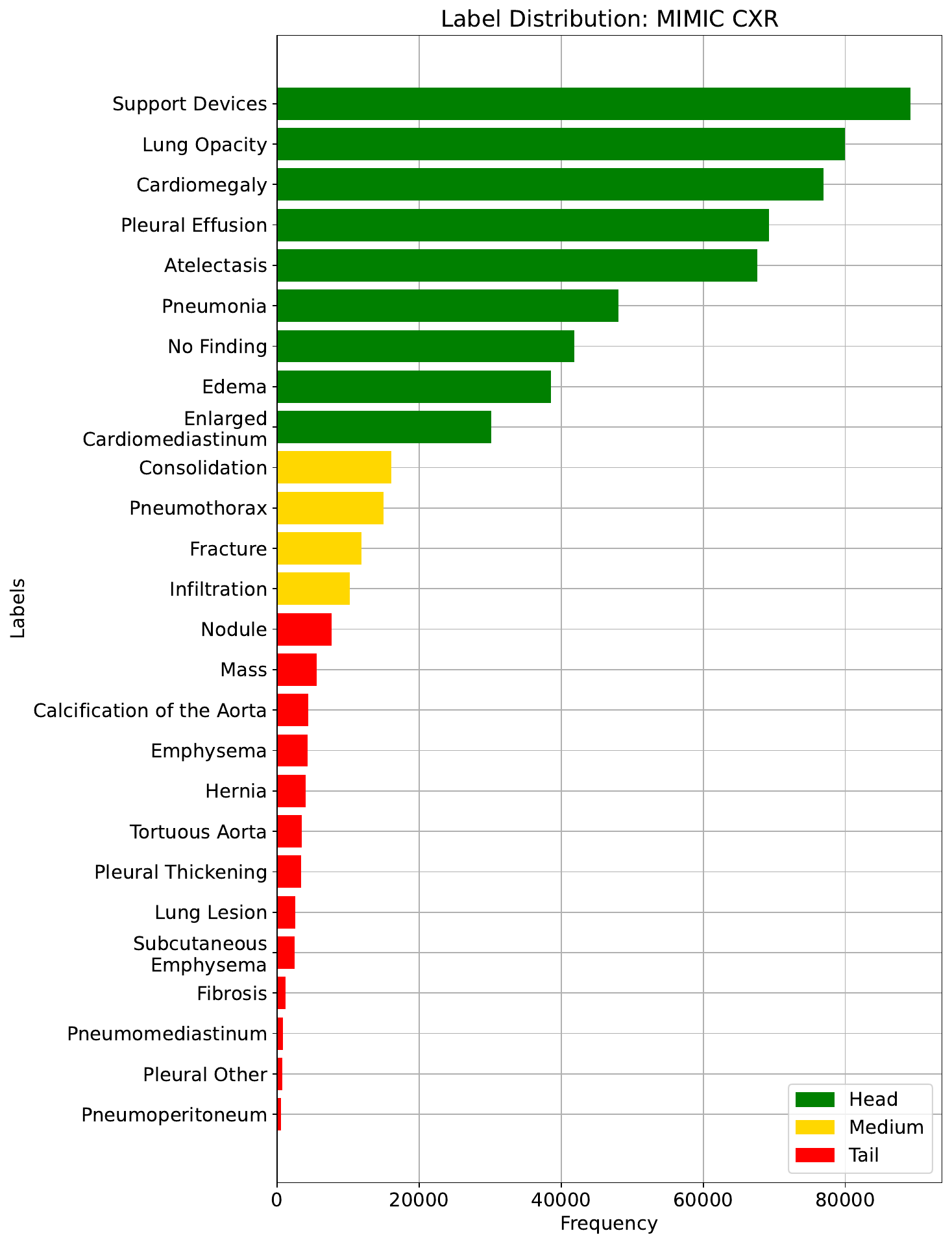}
    \vspace{-2mm}
    \caption{Distribution of the 26 labels in the MIMIC~CXR~\cite{johnson2019mimic} dataset. Head category labels contain $>$~30000 positive samples, medium category labels contain 10000~-~30000 positive samples, while labels with $<$~10000 positive samples serve as the tail labels.}
    \vspace{-4mm}
    \label{fig:mimic_distribution}
\end{figure}

This long-tail distribution of labels (refer Figure \ref{fig:mimic_distribution}) poses a keen challenge for deep learning researchers developing models for CXRs \cite{lakhani2017deep, tang2020automated}. The difficulty is exacerbated by the multi-label nature of the task, i.e. many findings can co-occur independently on the chest X-ray. The models must learn to classify N findings simultaneously, and they must do this not only for the frequent findings (referred to as ``head'' labels) but also for the rarest classes (i.e. ``tail'' labels). Additionally, since CXR datasets are grayscale, models do not have access to the important resource of colour information to discriminate between classes.

Researchers have addressed the multi-label nature of chest X-ray tasks \cite{wang2023bb, chen2020label}, but have not studied the performance of the model explicitly on the tail labels, hence giving no clarity whether the model actually learns to correctly predict the rare labels. Other works analyzing the performance of models on the long-tail of medical imaging datasets study it as a multi-class classification problem \cite{yang2022proco}. 

Recent research \cite{holste2022long} has addressed the multi-label long-tail distribution problem, and the authors have released a dataset by adding new labels to the MIMIC~CXR~\cite{johnson2019mimic} and NIH~CXR~\cite{summers2019nih, wang2017chestx} datasets which have a very low frequency. They analyze the performance of these models on various baseline methods \cite{lin2017focal, zhang2017mixup}.

Researchers have studied the benefits of transfer learning and domain specific pretraining on the general performance of CXR models \cite{sellergren2022simplified, matsoukas2022makes}. In this paper, we show that the generalist pretraining (i.e. on ImageNet \cite{deng2009imagenet}) followed by domain-specific pretraining (i.e. on CXRs) \cite{reed2022self, verma2022can} particularly benefits the long-tail labels, even though labels for the same may not have been available during the pretraining stage. While we do not intend to dispute the potential advantages of more sophisticated methods like novel loss functions \cite{wu2020distribution}, resampling \cite{galdran2021balanced} etc., these often require extensive hyperparameter tuning, and will almost always require retraining the models from scratch whenever a new label is added. On the contrary, since our method relies on using initializations, this method can be extended to new classes by finetuning for a few epochs. Further, our method is not exclusive and may be combined with other techniques for further performance improvements. 
\section{Experimental Setup}
\label{sec:experiments}
\mypara{Datasets.}
We have used four open source datasets for our experiments. Out of these, NIH~CXR~\cite{summers2019nih, wang2017chestx}, PadChest~\cite{bustos2020padchest} and CheXpert~\cite{irvin2019chexpert} were used for pretraining (ref. Section~\ref{sec:supp_datasets} for details) and were split into an 80-10-10 ratio for training, validation, and testing respectively. 
The MIMIC~CXR~\cite{johnson2019mimic} dataset containing the expanded set of labels~\cite{holste2022long} was used for experiments in this paper. We split the subset released for the competition \cite{holstecxr} into train and validation splits (90-10 ratio respectively), and used it in the finetuning stage. A hidden subset was reserved by the organizers as a test set. We report the results on our validation set and the hidden test set in Table~\ref{tab:all_results}. Only MIMIC~CXR was chosen as the dataset for evaluation to simulate the hidden test set.
The splits were created at the patient level, and all available CXR views were considered. The labels, their counts and their categorization in head, medium and tail classes for MIMIC~CXR dataset are shown in Figure \ref{fig:mimic_distribution}. 

\mypara{Method.}
We initialize all our models using ImageNet \cite{deng2009imagenet} weights which are publicly available in the Torchvision \cite{marcel2010torchvision} library. The dimension of the final fully connected layer is equal to the number of labels in the union of label sets of all datasets described above, i.e. 26 labels. 

We establish our baseline by directly finetuning the model on the MIMIC CXR train set. In our method, we perform the in-domain training by training jointly on NIH CXR, PadChest and CheXpert datasets, which serves as the chest X-ray (CXR) initialization. Since these three datasets do not cover all the labels for which the fully connected layer was initialized, we only perform backpropagation for labels that are present (1) or absent (0) as per the annotations. Post the first stage of pretraining, the model is finetuned end-to-end on MIMIC CXR dataset like the baseline method.

Finally, we average the sigmoid scores produced by both models (referred to as ``Averaged'' in Table \ref{tab:all_results}). Here, the data for each phase of training is exactly the same, but the model architecture changes as indicated.

\mypara{Training Details.}
We resize all images to 448 $\times$ 448 resolution \cite{sabottke2020effect} contrary to the standard convention of resizing to 224 $\times$ 224 resolution in previous literature. This is because many small features may get lost on reducing the image resolution. In doing this, we tradeoff a reduced batch size and increased training time for significantly better results. We use DenseNet161 and ResNext101 as our backbones for different experiments with a fully connected layer on top. All our models are finetuned with the Binary Cross Entropy loss function with sigmoid activation functions in the final layer. The models are trained for 20 epochs, with a learning rate of $10^{-4}$, decayed by half every 5 epochs, and the Adam optimizer. We use the same hyperparameters to perform the CXR pretraining but train only for 10 epochs.

\mypara{Leveraging Synthetic Data.}
With the advent of generative modelling tools, we are armed with the power to mitigate the problem of having a long-tail by supplementing the rare labels. We run a small experiment to evaluate whether we can use RoentGen \cite{chambon2022roentgen}, a prompt-based Stable Diffusion model to generate realistic chest X-rays and augment the training data. We generate about 5000 X-rays for training  containing at least one pathology from the tail classes. We curate prompts for RoentGen by translating the Spanish reports available in the PadChest dataset to English, and also generate radiologist report-like prompts using ChatGPT \footnote{\href{https://chat.openai.com}{chat.openai.com}} by providing curated templates. Figure \ref{fig:Roentgen_samples} shows examples of prompts and the corresponding synthetic X-rays generated by RoentGen and verified by a radiologist.

\mypara{Evaluation Metrics.}
We report the Average Precision (AP) \cite{holstecxr} score as the metric for all classes, and compute mean AP by macro-averaging the scores of over all classes. This metric is preferred over precision, recall and F1 score since it is threshold agnostic. The theoretical baseline for AP of each class is the prevalence, which is usually very low in chest X-ray interpretation tasks. Hence, this metric holds the edge over AUROC, whose theoretical baseline is 0.5, and is often inflated for all labels, making it difficult to analyse and compare the performance of models.

\section{Results and Discussion}
\label{sec:results}
\begin{table*}[t]
\small
\tabcolsep=0.12cm
\centering
\vspace{-2mm}
\begin{tabular}{@{}cclccccccccc@{}}
\toprule
\multicolumn{1}{l}{} &
  \multicolumn{1}{l}{} &
  \textbf{Label} &
  \textbf{Prevalence} &
  \multicolumn{4}{c}{\textbf{ImageNet $\rightarrow$ MIMIC CXR}} &
  \multicolumn{4}{c}{\textbf{CXR $\rightarrow$ MIMIC CXR}} \\
\multicolumn{1}{l}{} &
  \multicolumn{1}{l}{} &
  \multicolumn{1}{l}{} &
   &
  \textbf{ResNeXt} &
  \textbf{DenseNet} &
  \textbf{Averaged} &
  \textbf{Test Set} &
  \textbf{ResNeXt} &
  \textbf{DenseNet }&
  \textbf{Averaged} &
  \textbf{Test Set} \\ \midrule
 \multirow{9}{*}{\rotatebox[origin=c]{90}{\textbf{Head}}} & 1                           & Support Devices            & 0.1403 & 0.8606 & 0.8392 & 0.8686 & 0.8944 & 0.8947 & 0.8839 & \textbf{0.9251} & \textcolor{forestgreen}{0.9095} \\
                    &      2                        & Lung Opacity               & 0.1258 & 0.5075 & 0.5356 & \textbf{0.5901} & 0.5827 & 0.5791 & 0.5799 & 0.5813 & \textcolor{forestgreen}{0.5982} \\
                   &         3                     & Cardiomegaly               & 0.1210  & 0.5710  & 0.5663 & 0.6151 & 0.6183 & 0.5963 & 0.6075 & \textbf{0.6448} & \textcolor{forestgreen}{0.6522} \\
                &        4                      & Pleural Effusion           & 0.1089 & 0.7617 & 0.7662 & 0.8081 & 0.7985 & 0.7817 & 0.7895 & \textbf{0.8169} & \textcolor{forestgreen}{0.8054} \\
                    &    4                          & Atelectasis                & 0.1064 & 0.5300   & 0.5265 & 0.5964 & 0.5877 & 0.5825 & 0.5719 & \textbf{0.6081} & \textcolor{forestgreen}{0.6099} \\
                   &           6                   & Pneumonia                  & 0.0757 & 0.2518 & 0.2489 & 0.2634 & 0.2668 & 0.2964 & 0.2994 & \textbf{0.3111} & \textcolor{forestgreen}{0.3079} \\
                    &      7                        & No Finding                 & 0.0659 & 0.4115 & 0.4191 & 0.4484 & 0.4515 & 0.4399 & 0.4402 & \textbf{0.4710}  & \textcolor{forestgreen}{0.4678} \\
                  &        8                      & Edema                      & 0.0607 & 0.4820  & 0.4999 & 0.5262 & 0.5263 & 0.5236 & 0.5309 & \textbf{0.5477} & \textcolor{forestgreen}{0.5568} \\
                   &     9                         & Enlarged Cardiomediastinum & 0.0474 & 0.1365 & 0.1398 & 0.1734 & 0.1750  & 0.1573 & 0.1561 & \textbf{0.1837} & \textcolor{forestgreen}{0.1846} \\
\hdashline
\multirow{4}{*}{\rotatebox[origin=c]{90}{\textbf{Medium}}}                   & 10       & Consolidation              & 0.0252 & 0.1684 & 0.1738 & 0.1984 & 0.1998 & 0.1955 & 0.2042 & \textbf{0.2248} & \textcolor{forestgreen}{0.2302} \\
                   &     11                         & Pneumothorax               & 0.0236 & 0.4169 & 0.4148 & 0.4914 & 0.4906 & 0.4788 & 0.4676 & \textbf{0.5387} & \textcolor{forestgreen}{0.5465} \\
                   &   12                           & Fracture                   & 0.0187 & 0.1667 & 0.1929 & 0.2321 & 0.2302 & 0.2169 & 0.2362 & \textbf{0.2633} & \textcolor{forestgreen}{0.2620}  \\
                   &   13                           & Infiltration               & 0.0161 & 0.0479 & 0.0513 & 0.0520  & 0.0528 & \textbf{0.0631} & 0.0619 & 0.0576 & \textcolor{forestgreen}{0.0576} \\
\hdashline
\multirow{8}{*}{\rotatebox[origin=c]{90}{\textbf{Tail}}}                   &  14        & Nodule                     & 0.0121 & 0.1150  & 0.1360  & 0.1716 & 0.1717 & 0.1925 & \textbf{0.1964} & 0.1932 & \textcolor{forestgreen}{0.1963} \\
                   &   15                           & Mass                       & 0.0087 & 0.1087 & 0.1296 & 0.1824 & 0.1862 & 0.1646 & 0.1773 & \textbf{0.2284} & \textcolor{forestgreen}{0.2269} \\
                   &   16                           & Emphysema                  & 0.0067 & 0.1231 & 0.1660  & 0.1386 & 0.1370  & 0.1334 & 0.1262 & \textbf{0.1856} & \textcolor{forestgreen}{0.1840}  \\
                   &   17                           & Hernia                     & 0.0064 & 0.3321 & 0.4074 & 0.4873 & 0.4915 & 0.4540  & 0.5060  & \textbf{0.5355} & \textcolor{forestgreen}{0.5384} \\
                   &   18                           & Pleural Thickening         & 0.0053 & 0.0539 & 0.0766 & 0.0858 & \textcolor{forestgreen}{0.0847} & 0.0875 & \textbf{0.1017} & 0.0812 & 0.0832 \\
                   &   19                           & Lung Lesion                & 0.0040  & 0.0198 & 0.0371 & 0.0280  & 0.0284 & 0.0416 & \textbf{0.0589} & 0.0306 & \textcolor{forestgreen}{0.0312} \\
                   &   20                           & Fibrosis                   & 0.0018 & 0.1188 & 0.1190  & 0.1163 & 0.1169 & 0.1265 & 0.1446 & \textbf{0.1515} & \textcolor{forestgreen}{0.1538} \\
                   &   21                           & Pleural Other              & 0.0011 & 0.0070  & 0.0179 & 0.0175 & \textcolor{forestgreen}{0.0177} & 0.0244 & \textbf{0.0389} & 0.0153 & 0.0157 \\
\hdashline
\multirow{5}{*}{\rotatebox[origin=c]{90}{\textbf{Tail (Tail-U)}}}                   & 22 & Calcification of the Aorta & 0.0069 & 0.1074 & 0.1164 & 0.0852 & 0.0847 & 0.1169 & \textbf{0.1380}  & 0.1077 & \textcolor{forestgreen}{0.1090}  \\
                   &    23                          & Tortuous Aorta             & 0.0055 & 0.0434 & 0.0417 & 0.0440  & 0.0443 & 0.0478 & 0.0501 & \textbf{0.0559} & \textcolor{forestgreen}{0.0555} \\
                   &   24                           & Subcutaneous Emphysema     & 0.0039 & 0.4830  & 0.4064 & 0.4440  & 0.4426 & 0.4533 & 0.4390  & \textbf{0.5268} & \textcolor{forestgreen}{0.5197} \\
                   &  25                            & Pneumomediastinum          & 0.0012 & 0.0930  & 0.0921 & 0.2214 & 0.2181 & 0.2642 & \textbf{0.2918} & 0.2858 & \textcolor{forestgreen}{0.2840}  \\
                   &   26                           & Pneumoperitoneum           & 0.0009 & 0.1495 & 0.0927 & 0.1174 & 0.1153 & 0.1518 & 0.1441 & \textbf{0.2388} & \textcolor{forestgreen}{0.2374} \\ \midrule
\multicolumn{1}{l}{} & \multicolumn{1}{l}{}         & Mean (All Labels)          & 0.0385 & 0.2718 & 0.2774 & 0.3078 & 0.3082 & 0.3102 & 0.3170  & \textbf{0.3389} & \textcolor{forestgreen}{0.3394} \\
\multicolumn{1}{l}{} & \multicolumn{1}{l}{}         & Mean (Head)                & 0.0947 & 0.5014 & 0.5046 & 0.5433 & 0.5446 & 0.5391 & 0.5399 & \textbf{0.5655} & \textcolor{forestgreen}{0.5658} \\
\multicolumn{1}{l}{} & \multicolumn{1}{l}{}         & Mean (Medium)              & 0.0209 & 0.2000    & 0.2082 & 0.2435 & 0.2434 & 0.2386 & 0.2425 & \textbf{0.2711} & \textcolor{forestgreen}{0.2741} \\
\multicolumn{1}{l}{} & \multicolumn{1}{l}{}         & Mean (Tail)                & 0.0050  & 0.1350  & 0.1415 & 0.1646 & 0.1645 & 0.1737 & 0.1856 & \textbf{0.2028} & \textcolor{forestgreen}{0.2027} \\
\multicolumn{1}{l}{} & \multicolumn{1}{l}{}         & Mean (Tail-U)            & 0.0037 & 0.1753 & 0.1499 & 0.1824 & 0.1810  & 0.1806 & 0.1795 & \textbf{0.2430}  & \textcolor{forestgreen}{0.2411} \\ \bottomrule
\end{tabular}
\caption{\textbf{Average Precision scores} of all the models pretrained with different strategies reported on the validation set. We also report the results on the competition's hidden test set for the ``Averaged'' model for both pretraining strategies. The numbers in \textcolor{forestgreen}{green} are the best test set results. ``Averaged'' corresponds to results obtained by averaging the sigmoid scores of both models. All evaluation splits are from the MIMIC~CXR~\cite{johnson2019mimic} dataset. Tail-U represents the score on the tail labels unique to MIMIC CXR.}
\label{tab:all_results}
\end{table*}

We report the results for each label, as well as the mean AP for all our models in Table \ref{tab:all_results}. The results for all models are reported on the validation set prepared by us, and on the hidden test set from the competition~\cite{holstecxr} for the best models from each pretraining strategy (i.e. ``Averaged'').

\mypara{Overall Results.}
The strategy of dual pretraining on ImageNet followed by Chest X-ray datasets is clearly the winner when we look at the mAP over all 26 labels. 25/26 results on chained pretraining strategy are better than the corresponding values for pretraining just on ImageNet, with absolute gains varying between 1-8\% AP points, and mean gains ranging between 3-4\% AP points between corresponding models. We also note that DenseNet tends to consistently perform slightly better than the ResNeXt architecture (more than 0.5\% mAP points better on the absolute scale), outperforming even the overall best model in 6/25 labels (while ResNeXt achieves this only for 1 label). However, we have not studied the reason behind this trend. It is also interesting to note that the models trained with chained pretraining beat the corresponding baseline on all labels barring a few exceptions (ex. Subcutaneous Emphysema for ResNeXt, Emphysema for DenseNet, and Pleural Thickening, Pleural Other and Lung Opacity for ``Averaged''). We see a similar trend on the test set results, where the chained pretraining strategy performs better for 24/26 labels, and is very close for the 2 labels where it is worse.

Table \ref{tab:all_results} shows that the performance gains are distributed across all the categories, with the minimum advantage observed in medium category labels (i.e. about 2\% absolute mAP gain, from 0.5433 to 0.5655 on the validation set).

\mypara{Results on the Tail Labels.}
When evaluating the models by computing macro-averaged mAP, the results may be strongly biased results to major performance improvements on the head labels due to the abundance of data for them \cite{zhang2023deep}. A fine-grained analysis of the performance of only the tail labels yields that our strategy of additional pretraining on CXR datasets wins again, with the advantage about 4.5\% AP points. It is interesting to note that while the ImageNet initalization performs better for 2 of the tail labels, it's advantage is negligible compared to cases where the chained pretraining is better.

\mypara{Results on the Novel Tail classes.}
When we analyse the distribution of labels in Figure \ref{fig:mimic_distribution}, we notice that some of the tail labels (ex. Nodule, Mass, Fibrosis etc.) of MIMIC CXR may be supplemented by samples from other datasets (ref. Figure~\ref{fig:dataset_distribution}). Hence, we also evaluate the model performance on tail labels of the MIMIC CXR dataset that are unique to it. We refer to these labels as ``Tail-U'' in Table \ref{tab:all_results}. The performance on these labels is a true representation of the merit of our method to improve performance on the long-tail labels.

We observe that the chained pretraining strategy is strong for all 5 labels that are unique to MIMIC~CXR. However, the advantage is not as consistent. The gain is only 0.5\% points for ResNeXt, but it stretches to almost 3\% points for DenseNet.This is a significant boost given the extremely low prevalence of these labels. The most likely reason for this is the relatively strong performance of the baseline with the ResNeXt architecture on the Tail-U labels; it is almost as strong as the performance of DenseNet on the chained pretraining method.

\mypara{Contribution of Model Averaging.}
Averaging the scores predicted by multiple models gives a tangible performance increase, as noted in Table \ref{tab:all_results}. Particularly, we see a 6\% absolute increase in mAP over our best individual model for the unique tail labels (0.1806 vs 0.2430). This is because individual models are likely to be more erroneous for the labels where they have seen the least amount of data (and hence the lowest AP scores), and averaging predictions should help improve the performance. Although we demonstrate results by averaging over just the model architectures, the same can be extended to typical data ensembles as well. 

The proof of merit of this method lies in the fact that it demonstrates the best results for 19/26 labels on the validation set (we don't evaluate other models on the test set, so no comments can be made for that). Further, this method triumphs in all category-wise evaluations.

\begin{figure}
    \centering
    
    \begin{subfigure}{0.46\textwidth}
        \centering
        \includegraphics[width=\linewidth]{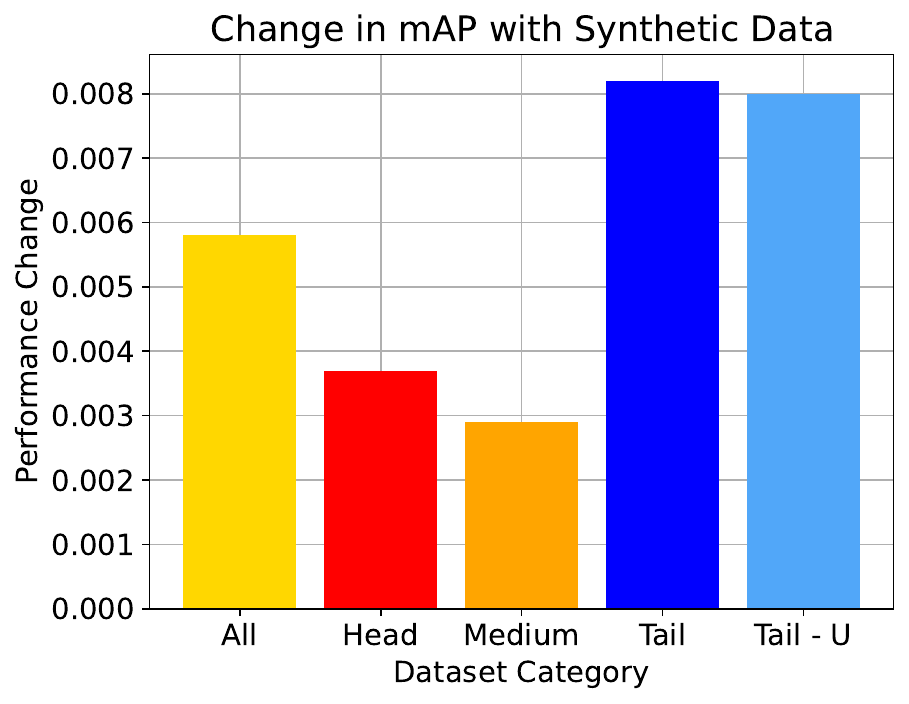}
    \end{subfigure}
    \vspace{-4mm}
    \caption{Increase in performance for each dataset category by adding 5000 synthetic data samples from tail classes.}
    \label{fig:delta}
\end{figure}

\begin{figure}[t]
    \centering
    \begin{subfigure}{0.225\textwidth}
        \centering
        \includegraphics[width=\linewidth]{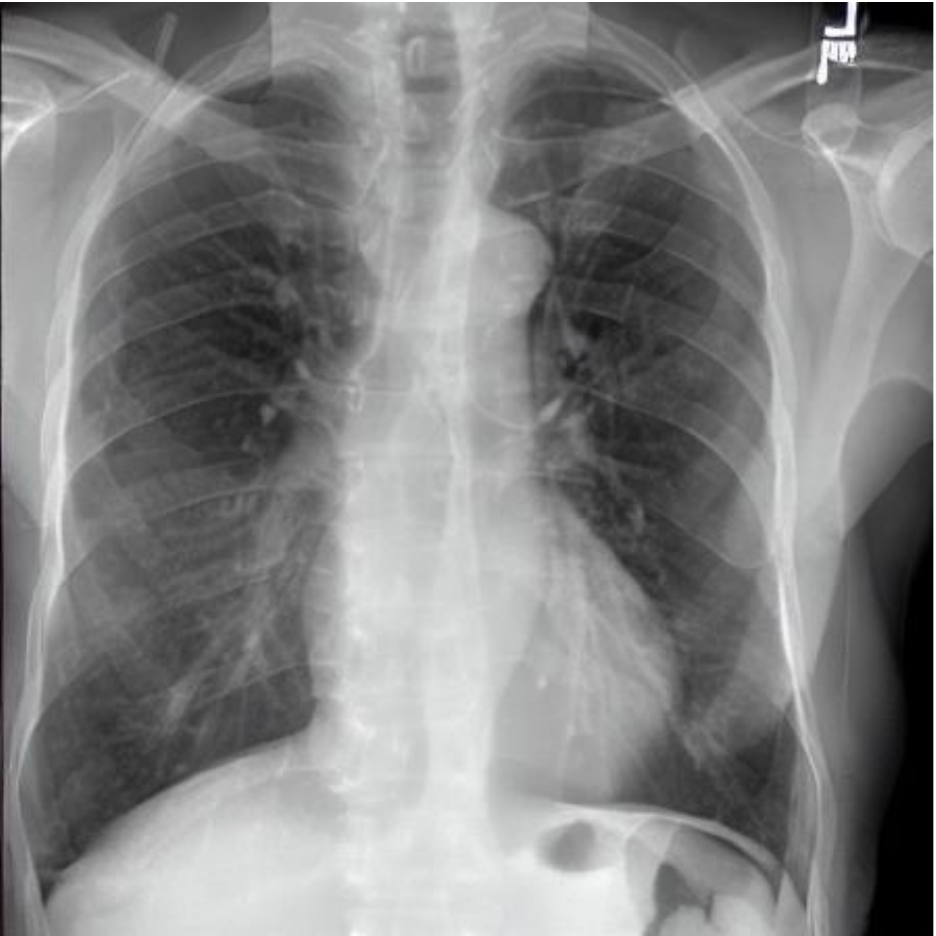}
        \caption{\underline{Pleural thickening} with calcifications is noted, which may be indicative of asbestos exposure or previous infection, as well as \underline{aortic calcification.}}
        \label{fig:ro1}
    \end{subfigure}
    \vspace{3mm}
    \hfill
    \begin{subfigure}{0.225\textwidth}
        \centering
        \includegraphics[width=\linewidth]{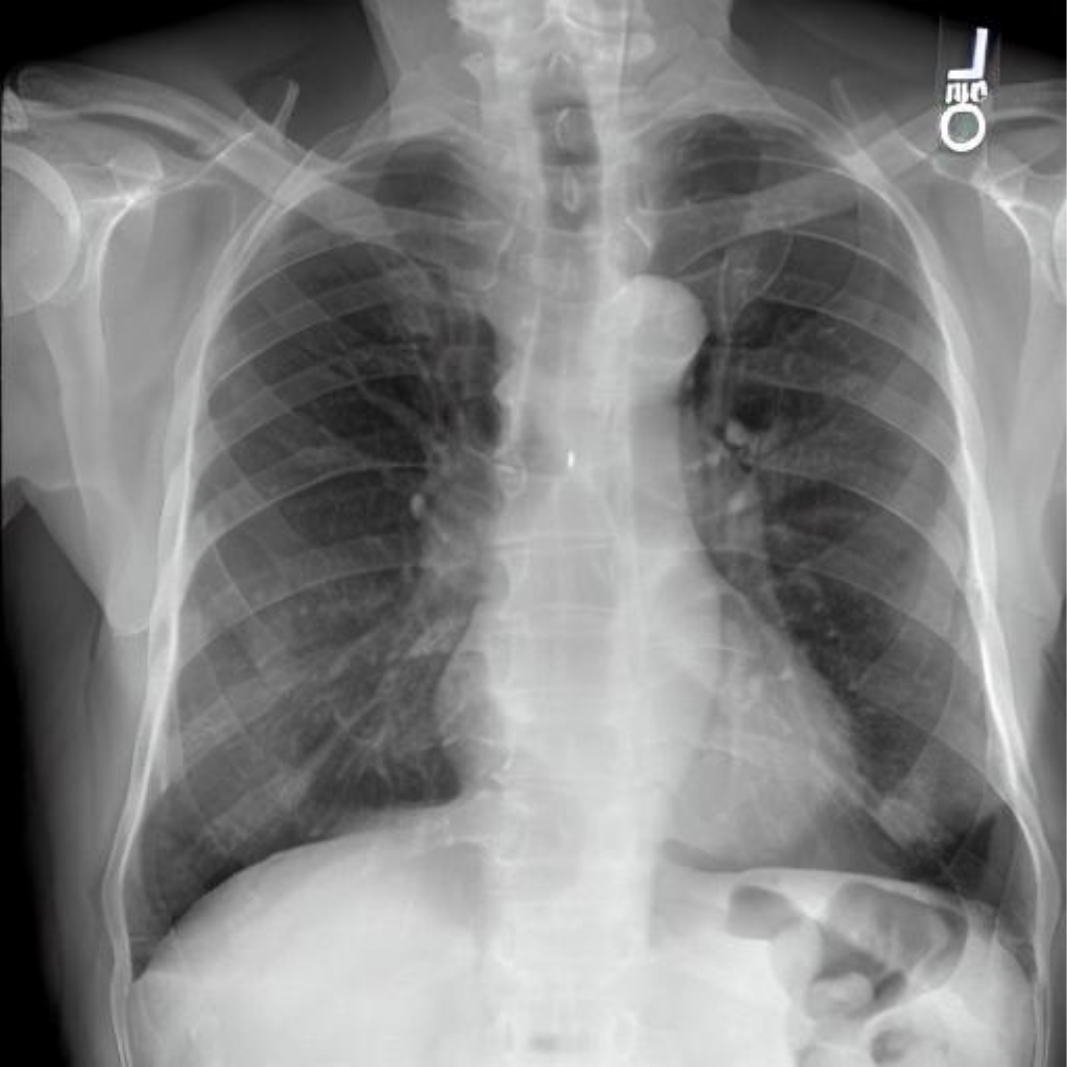}
        \caption{A \underline{thickened pleura} is noted, which may be indicative of previous infection, as well as \underline{calcification of the aorta}. Rest looks normal.}
        \label{fig:ro2}
    \end{subfigure}
    \vspace{3mm}
    \begin{subfigure}{0.225\textwidth}
        \centering
        \includegraphics[width=\linewidth]{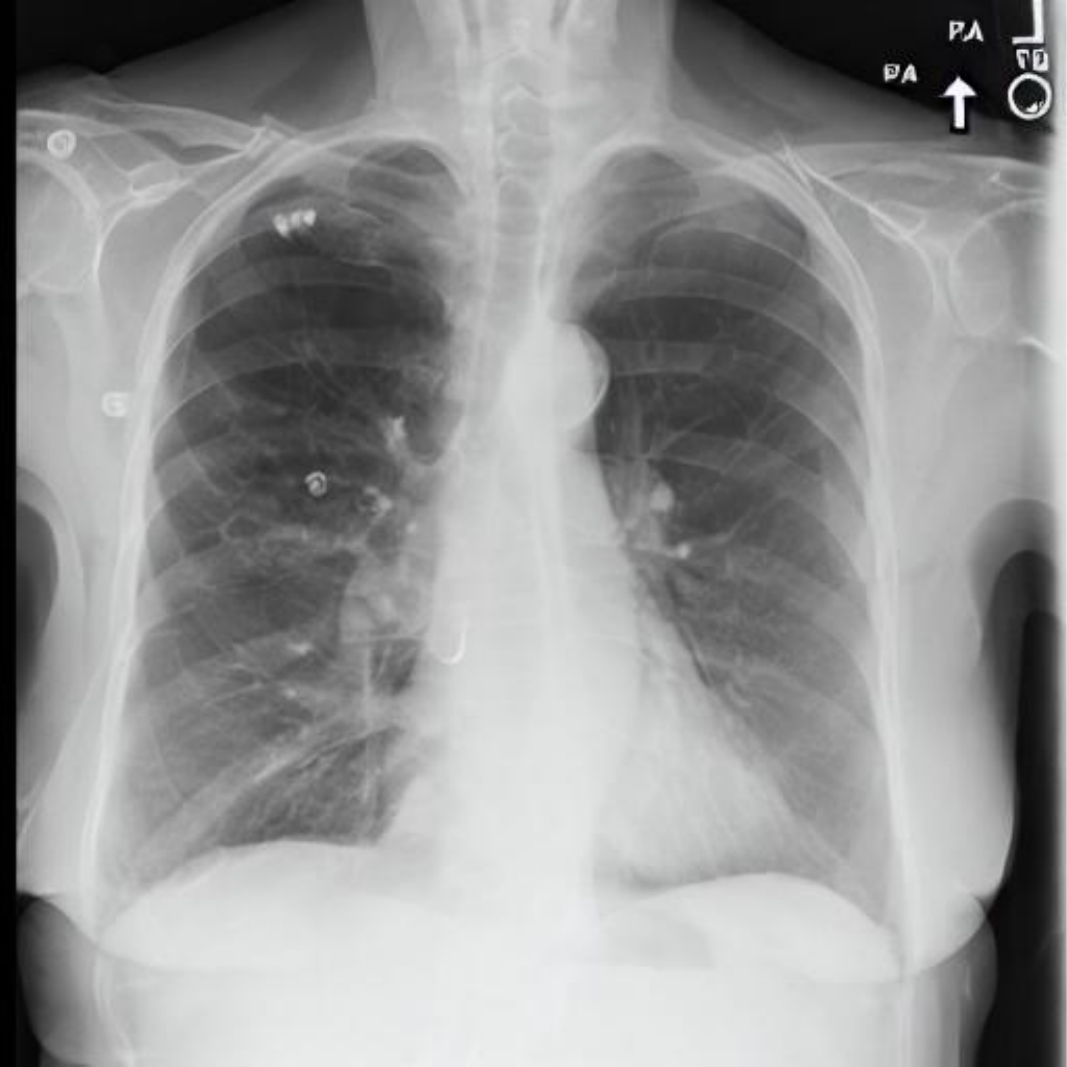}
        \caption{Mark signs of chronic obstructive pulmonary disease, \underline{emphysema}. Nodule of high density projecting in the middle right lung field, measuring one centimeter, suggesting differential diagnosis of isolated ossified granuloma.}
        \label{fig:ro3}
    \end{subfigure}
    \hfill
    \begin{subfigure}{0.225\textwidth}
        \centering
        \includegraphics[width=\linewidth]{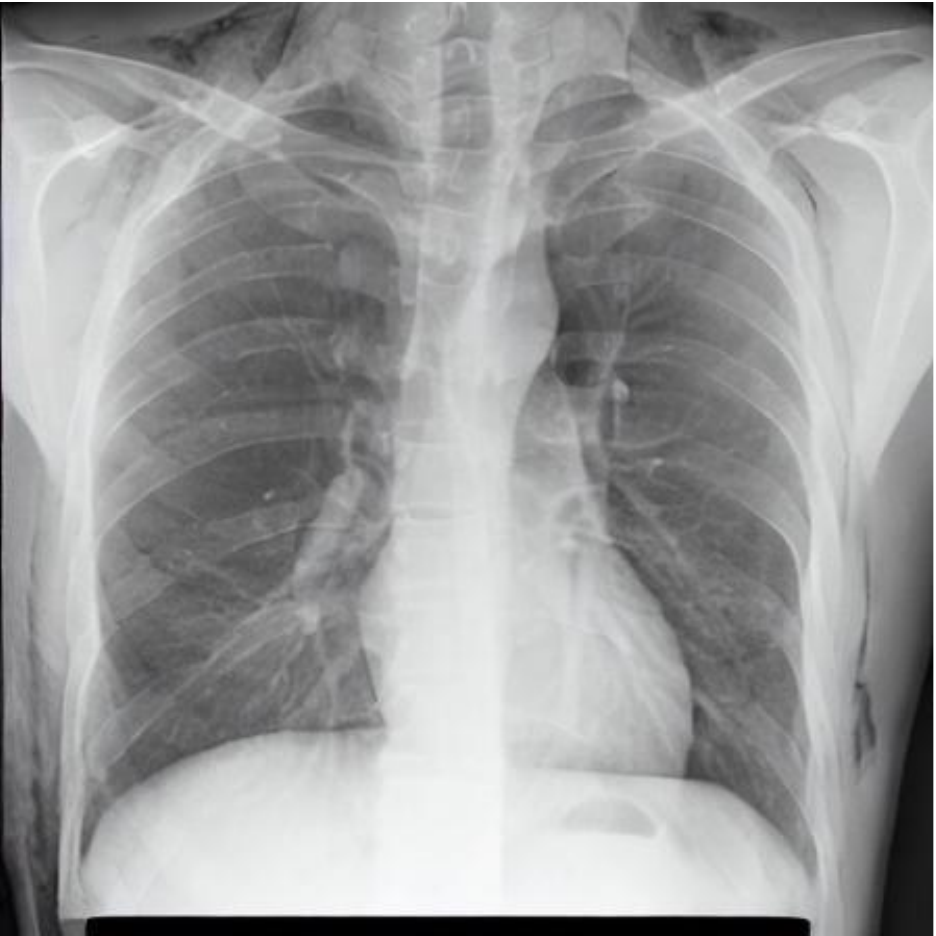}
        \caption{\underline{Subcutaneous emphysema} is seen in the subcutaneous tissues of the chest, extending from a \underline{pneumomediastinum}. The underlying cause should be explored, such as a possible esophageal perforation or trauma}
        \label{fig:ro4}
    \end{subfigure}
    \begin{subfigure}{0.225\textwidth}
        \centering
        \includegraphics[width=\linewidth]{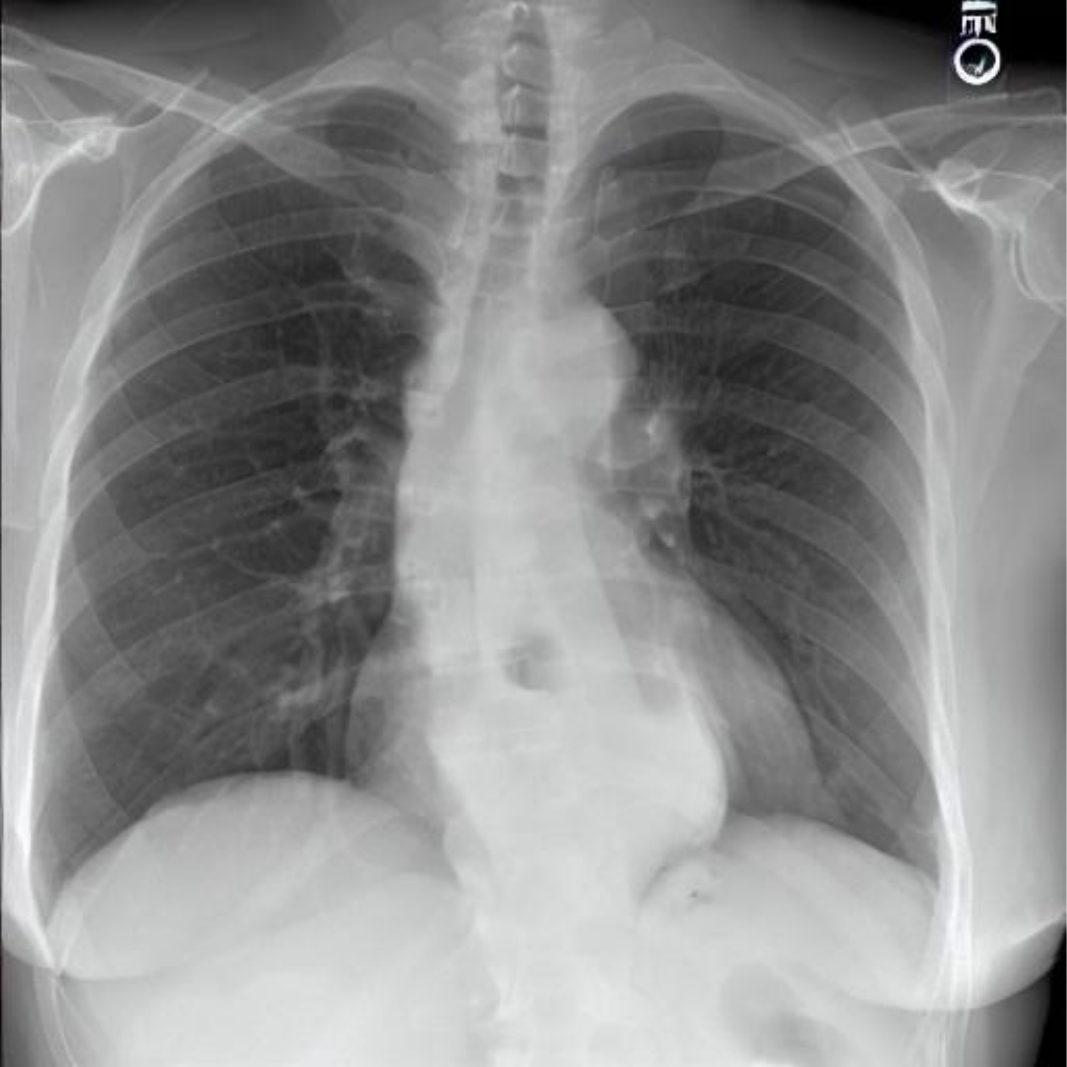}
        \caption{Inferior mediastinal \underline{mass} present with gastric fundus findings compatible with large hiatal \underline{hernia}. Aortic atheromatosis, degenerative changes in the dorsal column, and dorsolumbar scoliosis.}
        \label{fig:ro5}
    \end{subfigure}
    \hfill
    \begin{subfigure}{0.225\textwidth}
        \centering
        \includegraphics[width=\linewidth]{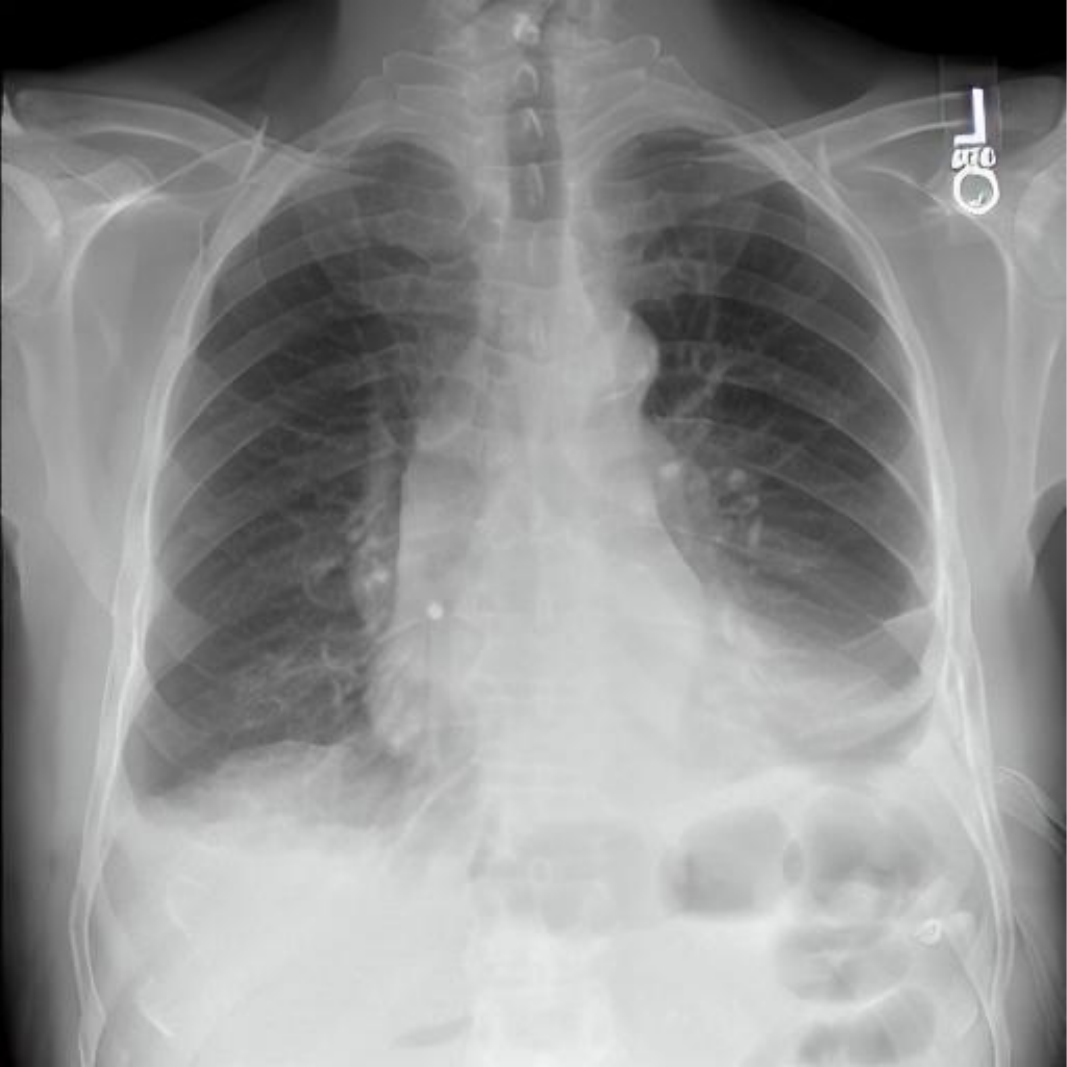}
        \caption{\underline{Infiltration} of the lower lobes in association with \underline{thickened pleura} and left-sided \underline{pleural effusion} suggests congestive heart failure with bilateral basal involvement. No other abnormalities.}
        \label{fig:ro6}
    \end{subfigure}
    \caption{Sample captions and RoentGen generated images which contain atleast one tail label from MIMIC~CXR~\cite{johnson2019mimic}. The phrases associated with labels are underlined.}
    \vspace{-3mm}
    \label{fig:Roentgen_samples}
\end{figure}

\mypara{Contribution of Synthetic Data.}
Figure \ref{fig:delta} shows the increase in performance of the model over Table \ref{tab:all_results}. This model was jointly finetuned on MIMIC CXR and a small synthetic dataset (refer Figure \ref{fig:Roentgen_samples} for samples). Although the size of the synthetic dataset was too small for it to demonstrate an appreciable performance delta, these results show the promise of leveraging synthetic data to overcome the challenge of training on rare classes at least to some extent, especially since the biggest change is seen in the performance of the tail classes.

\section{Conclusion}
\label{sec:conclusion}

We explore the effect of pretraining strategies for performance on chest X-ray tasks. Specifically, we analyze the performance improvement on the labels that reside in the tail of the label distribution, i.e. labels that rarely occur in the dataset. As expected, we observe a general trend that the performance of the model drops sharply as we move from the head of the distribution to the tail. However, we see that simply by leveraging checkpoints from a pretrained model on the source dataset, which may be publicly available \cite{cohen2022torchxrayvision}, we may able to improve the performance on infrequent tail labels without needing to experiment with hyperparameters or novel optimization objectives. Future work may explore the potential of further performance improvements by combining the chained pretraining technique with other methods to maximize the performance of deep learning models.

\paragraph{Acknowledgements.}
This work is made possible by the generous support of the American people through the United States Agency for International Development (USAID). The contents are the responsibility of Wadhwani AI and do not necessarily reflect the views of USAID or the United States Government.

\newpage
\balance
{\small
\bibliographystyle{ieee_fullname}
\bibliography{longstrings,refs}
}

\appendix 
\section*{Supplementary Material}
\label{sec:supp}

\section{Pretraining Datasets}
\label{sec:supp_datasets}

\begin{figure}[H]
    \centering
    \includegraphics[width=\linewidth]{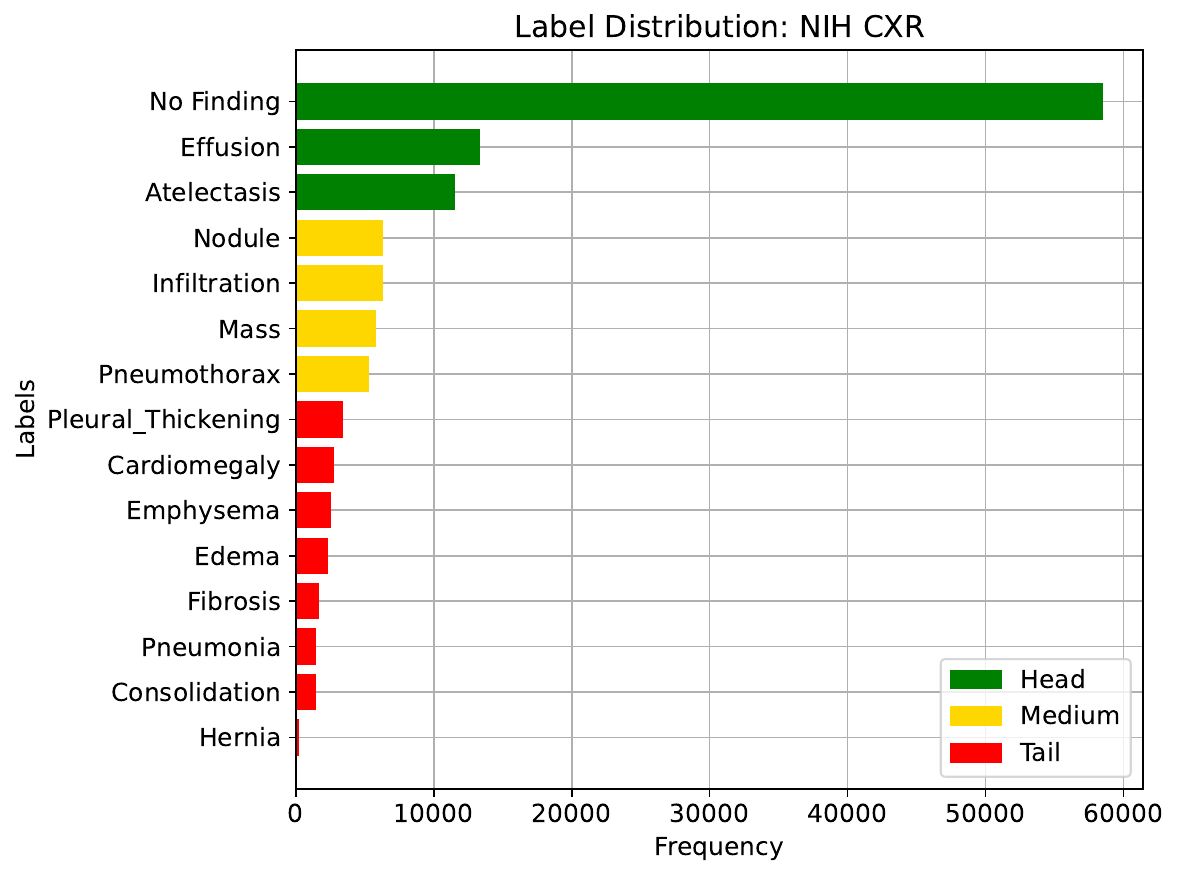} 
    \includegraphics[width=\linewidth]{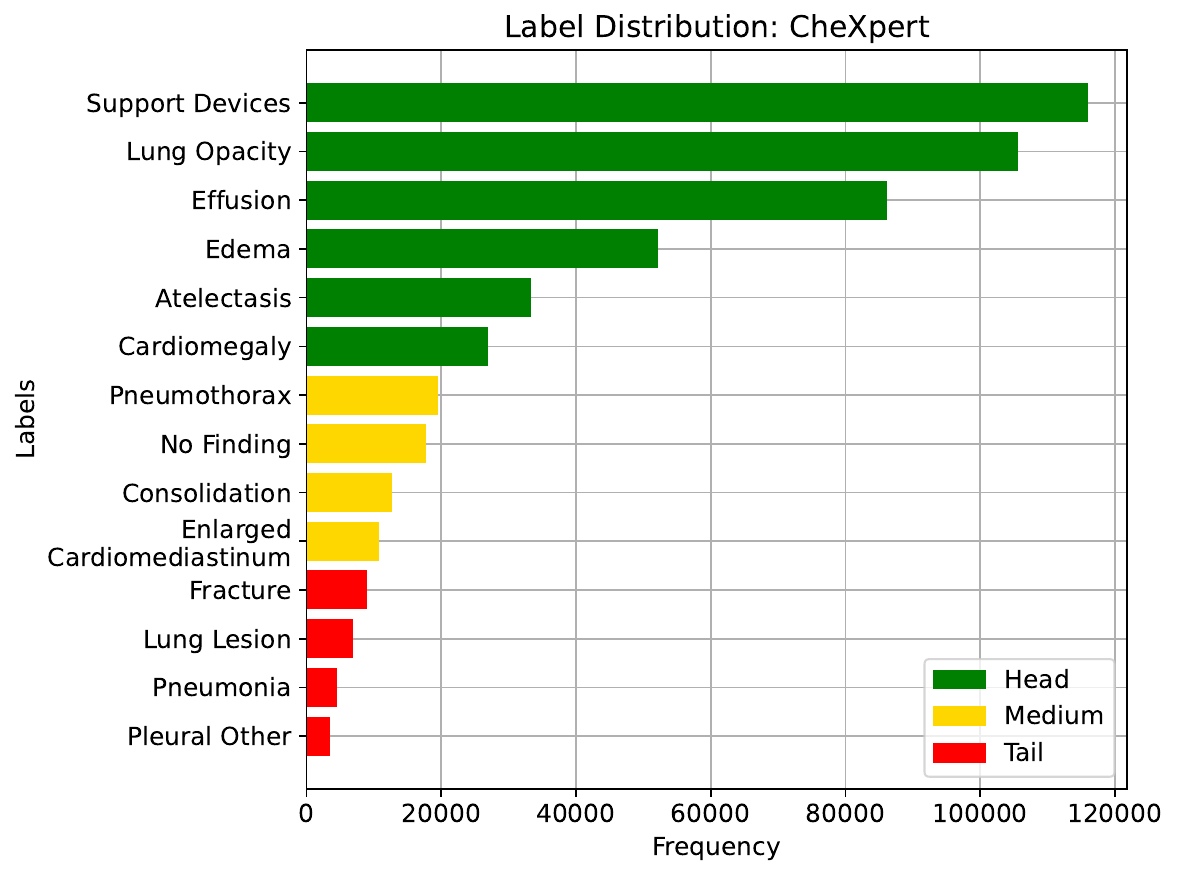}
    \includegraphics[width=\linewidth]{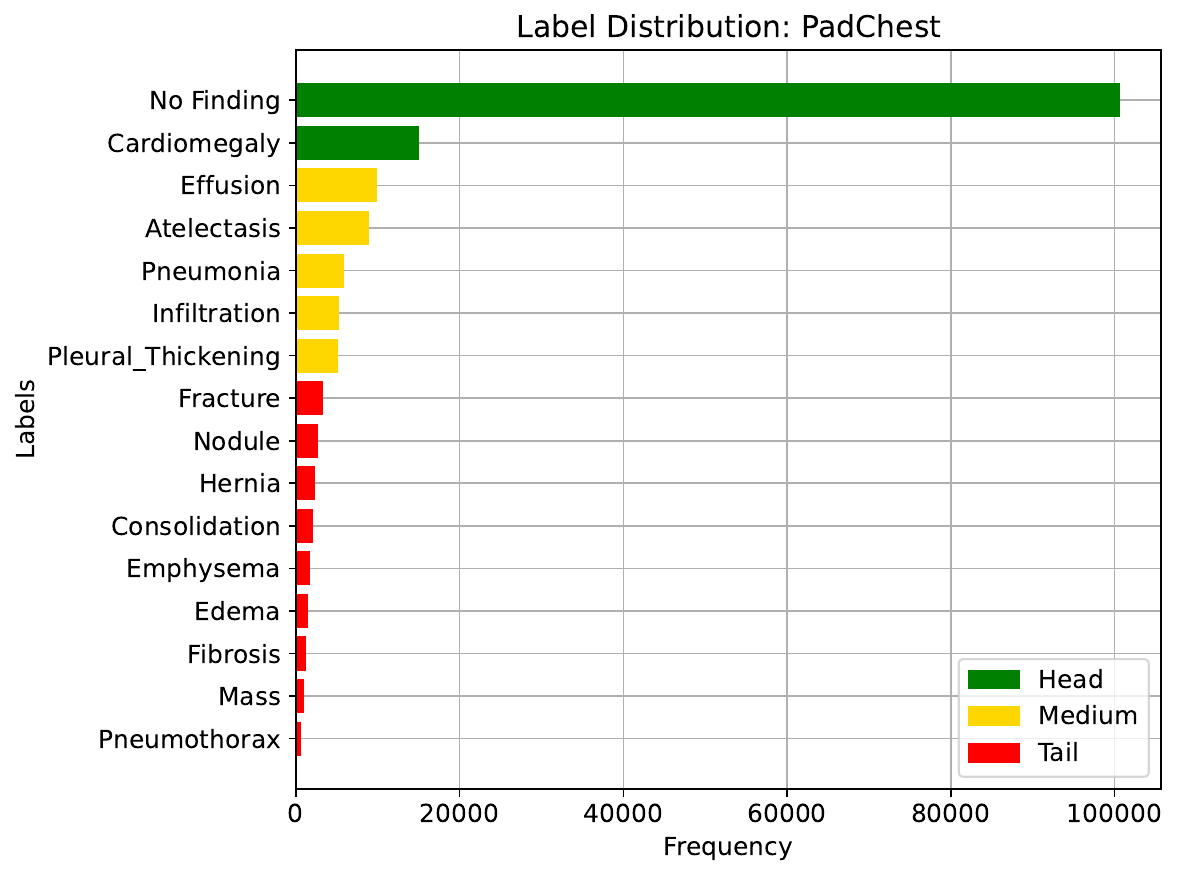}
    \vspace{-6mm}
    \caption{Distribution of labels in the NIH~CXR~\cite{summers2019nih,wang2017chestx} (\textbf{top}), CheXpert~\cite{irvin2019chexpert} (\textbf{middle}) and PadChest~\cite{bustos2020padchest} (\textbf{bottom}) datasets.}
    \label{fig:dataset_distribution}
\end{figure}

We show the distribution of labels in all three of our pretraining datasets - NIH~CXR~\cite{summers2019nih, wang2017chestx}, CheXpert~\cite{irvin2019chexpert} and PadChest~\cite{bustos2020padchest}. It is clear from Figure~\ref{fig:dataset_distribution} that each of these demonstrate a long tailed distribution to varying extents. 
Since the prevalence of labels is different for each dataset, and hence different thresholds were applied for this categorization. 

\paragraph{NIH~CXR dataset} \cite{summers2019nih, wang2017chestx} contains 15 labels. According to our categorization, it has 3 head category labels, each of which contain $>$~10000 positive samples. There are 4 medium category labels containing 5000~-~10000 positive samples, while the 8 tail category labels have $<$~5000 positive samples.

\paragraph{CheXpert dataset} \cite{irvin2019chexpert} contains 14 labels, with 6 head labels and 4 labels each in the medium and tail categories. The head category labels have more than 25000 positive samples, while the threshold for medium category samples is 10000 positives. All labels containing less than 10000 samples fall in the tail category. 

\paragraph{PadChest dataset} \cite{bustos2020padchest} contains 16 obtained after combination of various granular labels. It has only 2 head labels, with one of them - ``No Finding'' - being the most dominant. Further, it has 5 medium and 9 tail category labels.
We follow the same thresholds as NIH CXR for this categorization: $>$~10000 positive samples puts the label into the head category. Medium category labels contain 5000~-~10000 positive samples, and the tail category labels have $<$~5000 positive samples.

\end{document}